\documentclass[useAMS,usenatbib,twocolumn]{mnras}
\usepackage{graphicx}
\usepackage{amsbsy,amsmath}
\usepackage{graphics,graphicx}
\usepackage{subfloat}
\usepackage[compatibility=false]{caption}
\usepackage[skip=0.1pt]{subcaption}
\usepackage{ae,aecompl}
\usepackage[compact]{titlesec}
\usepackage{notoccite}
\usepackage{caption}
\usepackage{subcaption}
\usepackage{amsbsy,amsmath}
\usepackage{graphics,graphicx}
\usepackage{longtable}
\RequirePackage[latin1,utf8]{inputenc}
\usepackage{amsmath}
\usepackage{amsfonts}
\usepackage{amssymb}
\usepackage{makeidx}
\usepackage{float}
\restylefloat{figure}
\usepackage{stfloats}
\usepackage{epsfig}
\usepackage{graphicx}
\usepackage{caption}
\usepackage{color}
\usepackage{tabularx}
\usepackage{natbib}
\usepackage{multicol}
\usepackage{gensymb}
\usepackage[toc]{appendix}
\usepackage{breqn}
\usepackage{soul}
\setstcolor{red}
\usepackage{chngcntr}
\title{MOSS I: Double radio relics in the Saraswati supercluster}
\date{Accepted 2021 October 14. Received 2021 October 13; in original form 2021 August 19}

\author[Parekh et al.]{V. Parekh$^{1,2}$\thanks{E-mail: viralp@sarao.ac.za}, 
R. Kincaid$^{1}$,
K. Thorat$^{3}$, 
B. Hugo$^{1,2}$,
S. Sankhyayan$^{4,5}$,
R. Kale$^{4}$, \newauthor
N. Oozeer$^{6,2}$,
O. Smirnov$^{1,2}$,
I. Heywood$^{7,1,2}$,
S. Makhathini$^{8}$,
K. van der Heyden$^{9}$\\
$^{1}$Department of Physics and Electronics, Rhodes University, PO Box 94, Makhanda, 6140, South Africa\\
$^{2}$South African Radio Astronomy Observatory, 2 Fir Street, Black River Park, Observatory, Cape Town, 7925, South Africa\\
$^{3}$University of Pretoria, Lynnwood Rd, Hatfield, Pretoria, 0002, South Africa\\
$^{4}$National Centre for Radio Astrophysics, Savitribai Phule Pune University Campus, Ganeshkhind, Pune, Maharashtra 411007, India\\ 
$^{5}$Tartu Observatory, University of Tartu, Observatooriumi~1, 61602 T\~oravere, Estonia\\
$^{6}$African Institute for Mathematical Sciences, 6 Melrose Road, Muizenberg 7945, South Africa\\
$^{7}$Department of Physics, University of Oxford, Oxford OX1 2JD, UK\\
$^{8}$School of Physics, University of the Witwatersrand, 1 Jan Smuts Avenue, Johannesburg, 2000, Republic of South Africa\\
$^{9}$Department of Astronomy, R W James Building, University of Cape Town, Rondebosch, Cape Town, 7700, Republic of South Africa\\
}

\pubyear{2021}

\begin{document}
\label{firstpage}
\pagerange{\pageref{firstpage}--\pageref{lastpage}}
\maketitle

\begin{abstract}
\par Superclusters are the largest objects in the Universe, and they provide a unique opportunity to study how galaxy clusters are born at the junction of the cosmic web as well as the distribution of magnetic fields and relativistic particles beyond cluster volume. The field of radio astronomy is going through an exciting and important era of the Square Kilometer Array (SKA). We now have the most sensitive functional radio telescopes, such as the MeerKAT, which offers high angular resolution and sensitivity towards diffuse and faint radio sources. To study the radio environments around supercluster, we observed the (core part of) {\it Saraswati} supercluster with the MeerKAT. From our MeerKAT Observation of the {\it Saraswati} Supercluster (MOSS) project, the initial results of the pilot observations of two massive galaxy clusters, A2631 and ZwCl2341.1+0000, which are located around the dense central part of the {\it Saraswati} supercluster, were discussed.  In this paper, we describe the observations and data analysis details, including direction-dependent calibration. In particular, we focus on the ZwCl2341.1+0000 galaxy cluster, which hosts double radio relics and puzzling diffuse radio source in the filamentary network. We have imaged these double radio relics in our high resolution and sensitive L-band MeerKAT observation and a puzzling radio source, located between relics, in the low-resolution image. We also derived the spectra of double radio relics using MeerKAT and archival GMRT observations. A following papers will focus on the formation of radio relics and halo, as well as radio galaxy properties in a supercluster core environment.   

\end{abstract}

\section{Introduction}
\par Both high sensitivity observations and simulations show that the matter content in the Universe is not randomly distributed but resides in a complex structure called the `cosmic web' \citep{2015JCAP...01..036J,2015A&A...580A.119V,2014MNRAS.445.3706V,2006astro.ph..8019L,2005Natur.435..629S,1996Natur.380..603B}. Billions of stars, quasars, galaxies, and intergalactic gas make up the pattern of thin walls or filaments inside the Large-Scale Structure (LSS). These filaments are on the scales of a few tens to hundreds of Mpc and are surrounded by large voids. Furthermore, dense and massive galaxy clusters are born at the intersection of filaments. Hence, it is essential to study galaxy clusters in the `cosmic web' environment to understand how the Universe evolved and how matter flows within filaments and interacts with galaxy clusters. 
\par To date, mapping of the `cosmic web' was generally done using optical redshift surveys; finding densities in galaxy distributions was the only available tool to trace large scale filaments \citep{2014MNRAS.438.3465T}. Thankfully, technical enhancements in the observations of superclusters at other wavelengths such as radio band, have enhanced our knowledge of the cosmic web, its origin, evolution, physical condition, and properties \citep{2018MNRAS.479..776V, galaxies5010016, 2012MNRAS.423.2325A}. Further, there are many other critical questions that depend on the future high quality and sensitive multiwavelength observations to directly measure baryon density and its distribution in the cosmic web, as well as the total energy budget to drive the material inside filaments, magnetic field strength and overall growth of the structure. 
\par In the last two decades, galaxy clusters have been extensively studied in the radio bands. Low-frequency radio observations (at 1 GHz and below) have revealed the existence of Mpc size diffuse radio sources associated with the cluster formation process \citep[and references therein]{2012A&ARv..20...54F}. These extended diffuse radio sources further split into two classes -- radio halos and relics \citep[and references therein]{2019SSRv..215...16V}. These two classes of sources are similar in their radio properties, aside from their positions and morphologies. Giant radio halos are spherical and can generally be found at the cluster centre or core, while radio relics are elongated in one direction and located at the cluster peripheries. Formation mechanisms of these radio sources are an open debate; however, observations hint that these radio sources are generated due to the turbulence (radio halos) and shock structures (radio relics) activated by the cluster merging process. {Recently, low-frequency observations have also revealed another class of diffuse radio source, know as `radio-ridge' found between two galaxy clusters in low dense regions \citep{2019Sci...364..981G,2019A&A...630A..77B}.  The formation mechanism of this new class of radio source is not yet fully understood \citep{2020PhRvL.124e1101B}}.
\par In the present era, with the ability to conduct high sensitivity and angular resolution radio observations with next-generation telescopes, such as the MeerKAT \citep{2018ApJ...856..180C,2009IEEEP..97.1522J}, Australian Square Kilometer Array Pathfinder (ASKAP; \cite{2016PASA...33...42M}), and LOw Frequency ARray (LOFAR; \cite{2013A&A...556A...2V}), the time has arrived to utilise these telescopes to observe superclusters and filaments in order to detect shocks, particle acceleration and magnetic fields beyond cluster volume \citep{2020PhRvL.124e1101B,2019Sci...364..981G,2019A&A...630A..77B}. We have observed the {\it Saraswati} supercluster \citep{2017ApJ...844...25B} with the MeerKAT telescope to study the supercluster environment in radio frequency regime, characterise cluster merger shock and/or accretion shock related to relics, and mapping of diffuse radio sources with high dynamic range imaging. 
\par With the advancement in radio interferometric observations, there is an increase in demand to produce the most sensitive radio maps with the highest dynamic range \citep{2011A&A...527A.108S}. In order to do so, one has to deal with many observational and data calibration errors, which place constraints on achieving such sensitivity. These errors include antenna pointing inaccuracy, time and frequency-dependent primary beam variation across fields, stability of the ionosphere over which observation is conducted, unknown radio frequency interference (RFI) sources, thermal noise and gain variation of receivers, to name a few. These time and frequency varying errors can also be related to one or more source-specific directions, known as direction-dependent errors. Sometimes strong point sources in the primary beam can limit the sensitivity required for observations. This problem is more severe when studying diffuse radio sources in the vicinity of a strong radio source(s) which hampered the observation by generating artefacts. Primary calibration processes (even self-calibration) are not adequate enough to fully characterise the properties of these strong radio sources, hence these poorly deconvolved sources challenge us in the imaging of faint and diffuse radio sources. One common solution to this issue is known as ``source peeling'' \citep{2004SPIE.5489..817N}. The general principle here is to obtain source-specific calibration solutions towards the strong source (which we need to be peeled) location  or, in other words, apply direction-dependent calibration towards the particular source direction and solve the complex gain of that source and then subtracted it from the visibility. A number of software packages implement variations on this approach \citep[e.g.][]{2018MNRAS.478.2399K,Rioja_2018,2016ApJS..223....2V,2014ASInC..13..469I,2010A&A...524A..61N,2007ASPC..376..127M}, with the aim of producing the best scientific images. 
\subsection{{\it Saraswati} supercluster}
\par The {\it Saraswati} supercluster was recently discovered in the famous Stripe 82 region \citep{2011AJ....142....3H} of the Sloan Digital Sky Survey (SDSS; \citealt{2000AJ....120.1579Y}), at redshift $z$ $\sim$ 0.3 \citep{2017ApJ...844...25B}. The total mass and size of this supercluster are $\sim$ 2 $\times$ 10$^{16}$ $M_\odot$ and $\sim$ 200 Mpc, respectively, making it one of the largest observed structures in the Universe. This massive supercluster hosts 43 galaxy clusters at a mean redshift of $z$ $\sim$ 0.28 and has an average density contrast of $\delta = 1.62$. Furthermore, it is surrounded by a complex network of galaxy filaments and large voids. There are a total of 24 voids of the size $\sim$ 40-170 Mpc identified in the {vicinity of the} supercluster. {The {clean spectroscopic} redshifts are derived from the LEGACY \citep{2009ApJS..182..543A}, BOSS \citep{2013AJ....145...10D} and SOUTHERN\footnote{https://www.sdss.org/dr12/algorithms/legacy_special_target/\#southern} (LBS) programs of the SDSS-III DR12 database \citep{2015ApJS..219...12A}}. A total of 3016 galaxies have been found within {and around} the {\it Saraswati} supercluster region \citep{2017ApJ...844...25B}. 
\par The {\it Saraswati} supercluster's dense central (gravitationally bound) region extends to a radius of 20 Mpc and encompasses a mass of at least 4 $\times$ 10$^{15}$ $M_\odot$ (20\% of total supercluster mass) comprised of five massive galaxy clusters, including Abell 2631. These five massive clusters form the dense, bound core of the {\it Saraswati} supercluster. Abell 2631 is the most massive cluster located in the core of the {\it Saraswati} supercluster, and it is also extremely rich (richness class 3) and hot {($\sim$ 9.60 keV, \citealt{2009ApJS..182...12C})}.  The complex merging galaxy cluster ZwCl 2341.1+0000, which is situated at $\sim$ 45 Mpc from the core, is the second  massive cluster and is located in the trail of galaxies which makes up a southern filamentary network that is connected to the main central region. In the NRAO VLA Sky Survey (NVSS) data \citep{1998AJ....115.1693C}, it was found that ZwCl 2341.1+0000 hosts double giant diffuse radio relics connected by faint `radio-bridge'. The overall structure is indicative of the infall and merger dynamics of several galaxy groups/clusters during the initial phase of the cluster formation process \citep{2002NewA....7..249B}. Later on, in the Giant Meterwave Radio Telescope (GMRT) low-frequency observation of ZwCl 2341.1+000, double radio relics were detected, but there was no radio bridge or halo found between double relics (\citealt[][hereafter V09]{2009A&A...506.1083V}). However, there was marginalised faint radio emission detected between these two relics, along the filament network, in the Very Large Array (VLA) L-band observation (\citealt[][hereafter G10]{2010A&A...511L...5G}). However, due to lack of sensitivity in the VLA data, the nature of this faint emission could not be determined and could thus not be studied further. The estimated virial radii of both clusters are $\sim$ 1.93 Mpc and $\sim$ 1.62 Mpc, respectively, corresponding to the angular sizes of 8$'$ and 6.5$'$. 

In this paper, we present our MeerKAT and GMRT (archival data) analysis results of the MOSS. Firstly, we focus on the direction-dependent calibration technique for MeerKAT data for improving our images. We also show the point source subtraction method used to analyse diffuse radio sources. Both these techniques make use of new in-house software. 
\par As this section has provided a clear introduction to the research, the rest of the paper will be structured as follows: \S 2 gives details of the radio observation of the {\it Saraswati} supercluster; \S 3 describes the radio data reduction procedures; \S 4 presents our results; \S 5 gives the discussion and conclusion. In this paper, we have assumed $H_{0}$ = 70 km\,s$^{-1}$\,Mpc$^{-1}$, $\Omega_{\mathrm{M}}$ = 0.3 and $\Omega_{\Lambda}$ = 0.7. At redshift $z$ of 0.27, 1$''$ = 4.14 kpc, and luminosity distance $D_{\mathrm{L}}$ = 1376 Mpc.



\section{Radio observations}

\subsection{MeerKAT L-band observation}
\par MeerKAT is the precursor of the Square Kilometer Array (SKA) located in the Karoo desert of South Africa \citep{2018ApJ...856..180C,2009IEEEP..97.1522J}. We observed the central part of the {\it Saraswati} supercluster with the MeerKAT, which includes two pointings centred on each of the A2631 and ZwCl2341.1+0000 galaxy clusters. The separation between the two pointing is 1.5$^{\circ}$. MeerKAT observational details are provided in Table \ref{MK_observation}.  

\begin{table}
\caption{MeerKAT observations of the {\it Saraswati} supercluster.}
\label{MK_observation}
\begin{tabular}{lll}
\hline
\hline
&Observation date: &           2019-06-15  \\
&Observation time: &               00:41:27.8 \\
&Number of pointings: &            2 \\
&Phase centre of A2631 &  23h37m40.6s\\
&                       &  +0d16m36.0s\\
&Phase centre of ZwCl2341.1+0000 & 23h43m39.7s\\
&                                 & +0d19m51.0s\\
&Number of antennas: &             60\\
&Total observation time:&          14hrs  \\
&Central frequency: &              1283 MHz\\
&Total bandwidth:   &              900 MHz   \\
&Channel width: &                  208 kHz \\
&Total number of channels: &      4016 \\
&Dump time: &                      8s\\
&Cross products: &                  XX,XY,YX,YY\\ 
&Bandpass and flux calibrator:   & J1939--6342    \\
&Gain calibrator:   &              J2357--1125\\
\hline

\end{tabular}
\end{table}

\subsection{GMRT 325 MHz observation}
\par ZwCl 2341.1+0000 was observed with the legacy GMRT on 24 June 2011 and 20 July 2011 at 325 MHz. To the best of our knowledge, this archival data set was not published before. The pointing of antennas were chosen at RA: 23h43m45.012482s, DEC: +00d18m00.08115s. The total observation time was 22 hours. This observation was performed using 256 channels, 30 MHz bandwidth and four cross products. To fix the absolute amplitude scale and bandpass calibration, 3C286 and 3C48 calibrators were observed, along with the J2340+135 (RA: 23h40m33.232373s; DEC: +13d33m00.98106s) phase calibrator. 

\section{Data analysis}
\par We reduced both the MeerKAT and GMRT data sets using the \textbf{C}ontainerized \textbf{A}utomated \textbf{R}adio \textbf{A}stronomy \textbf{Cal}ibration (CARACal\footnote[1]{https://caracal.readthedocs.io/en/latest/}) pipeline \citep{2020arXiv200602955J}. The use of this pipeline is also mentioned in our previous work \cite{2020MNRAS.499..404P}. Briefly, the CARACal pipeline is written using the Stimela\footnote[2]{https://github.com/ratt-ru/Stimela} pipelining framework, which is python-based and uses containers (Docker and Singularity) specially developed for independent radio interferometry scripting, which allows users to use general radio astronomy software. This permits the combination of the appropriate radio astronomy software packages into a single pipeline. CARACal is an open-source and flexible pipeline that comes with pre-defined input settings for the reduction of data from MeerKAT and other telescopes like GMRT and JVLA in a variety of configuration files, which can be modified by the user as per their individual needs. We also made use of the Stimela package to carry out tasks such as excising discrete radio point sources (from {\it{uv}} domain) blended with extended emission and direction-dependent calibration. More details of both these methods are given in \S \ref{ddcal} and \S \ref{ptsub}.    


\subsection{MeerKAT data analysis}
\par {We used standard bandpass and gain calibration tasks of Common Astronomy Software Application (CASA) for cross-calibration inside the CARACal pipeline}, with several rounds of automatic RFI flagging (before and after calibration). We then used the AOFlagger \citep{offringa-2012-morph-rfi-algorithm} software with a custom strategy optimised for MeerKAT data to remove the RFI. We also applied the mask of known RFI-affected channels to the data. In order to fix the absolute scale of the flux calibrator, we used the local sky model of the calibrator derived using the Reynolds scale \citep{ReyATNF.352..508R}. The absolute flux density of the primary or flux calibrator J1939--6342 is $\sim$ 14.90 Jy at the rest frequency of 1.4 GHz. After bootstrapping the flux density scale from the flux calibrator, the flux density of the secondary calibrator J2357--1125 is 2.07 $\pm$ 0.01 at 1.28 GHz (at 1.4 GHz, its value is $\sim$ 1.8 Jy and after scaling to 1.28 GHz with a  spectral index value of -0.7, its expected value is $\sim$ {1.9 Jy}). Once the data were calibrated, we then split the two target sources from the multi-source measurement set. To reduce the data size and imaging time, we averaged every five channels in the target data. We then imaged ZwCl 2341.1+0000 and A2631 fields separately in the self-calibration rounds. In the pipeline, for direction-independent imaging and self-calibration tasks, we used the WSClean \citep{2014MNRAS.444..606O} and CubiCal \citep{2018MNRAS.478.2399K} software. We performed three phase-only self-calibration rounds to improve the antenna-based phase solutions. In the deconvolution procedure, we employed wide-band and multiscale cleaning options to better image the diffuse emission. We performed joint deconvolution on five sub-bands and generated a multi-frequency synthesis (MFS) cube (the bandwidth of each sub-band image is 170 MHz and has 164 channels). In joint deconvolution, we used the 2$^{nd}$ order polynomial (\texttt{nterms = 2}) in the spectral fitting. We made images with different Briggs robust weighting values to generate high- and low-resolution images. Further, we used these self-calibrated visibilities of ZwCl 2341.1+0000 and A2631 fields for direction-dependent calibration. 
\subsection{GMRT data analysis}
\par For the 325 MHz GMRT archival data reduction, we also used the CARACal pipeline. We used the default settings to perform the initial flux and gain calibrations. For RFI removal, we used the \texttt{tfcrop} and \texttt{rflag} options of CASA, accessible inside the pipeline. For the flux calibrator, we used the Perley-Butler 2010 (hereafter PB10) scale. This gave the flux density of the flux calibrator 3C286 to be $\sim$ 24.83 Jy. The phase calibrator for this observation was 2340+135 (RA:23h40m33.22s, Dec:+13d33m00.92s). Similar to the MeerKAT data, we performed the joint deconvolution on five sub-bands (each sub-band of 6 MHz bandwidth with 51 channels) with multiscale cleaning and generated an MFS image using the 2$^{nd}$ order polynomial fitting. We performed three rounds of phase and two rounds of amplitude and phase self-calibration. The goal of the present data analysis is to test the CARACal pipeline for the GMRT antennas and tune the pipeline for uGMRT wide-band data analysis for future use.

\subsection{MeerKAT and GMRT flux calibration accuracy}
\label{cal_ac}
\begin{figure}
\includegraphics[scale=0.4]{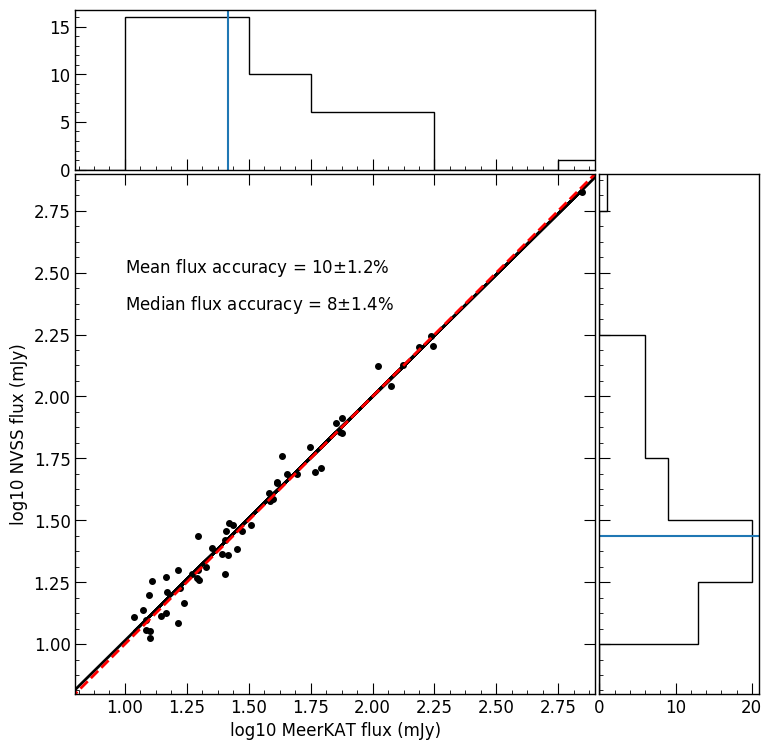}
\caption{Total flux density comparison between MeerKAT and NVSS for ZwCl2341.1+0000 and A2631 fields. The solid black line shows the best-fit relationship between two observations and dashed red line shows the one-to-one relationship. Each side histogram shows the flux density distribution for MeerKAT and NVSS, with the blue line indicating the median value.}
\label{NVSS_MeerKAT}
\end{figure}

In order to obtain the calibration accuracy, we compared the flux densities of unresolved and discrete point radio sources between MeerKAT and NVSS observations of the ZwCl2341.1+0000 and A2631 fields. In both images, we manually selected isolated bright point sources (of flux density $>$ 1 mJy) to avoid any source blending or overlapping. We made a list of these point sources and used the CASA's \texttt{imfit} task to fit the Gaussian and measure the flux densities from both MeerKAT and NVSS images. In this analysis, we used the MeerKAT primary beam \citep{2021MNRAS.502.2970A} corrected images. We scaled the 1283 MHz MeerKAT flux density values to 1400 MHz with a spectral index value of -0.7. We have shown our flux densities comparison result in Figure \ref{NVSS_MeerKAT}. We found that the overall (mean) flux density accuracy between MeerKAT and NVSS is $\sim$ 10\% and median flux density accuracy $\sim$ 8\%.. For GMRT observations, this error is also estimated to be 10\% \citep{2004ApJ...612..974C}. 

\subsection{Flux density error estimation and calibration scales}
\label{cal_scale}
\par In the procedure of uncertainty estimation on the total flux densities, two main factors should be evaluated. The first factor is the data calibration accuracy. From our estimation (\S\ref{cal_ac}), this calibration error was measured to be $\sim$ 10$\%$.  The second factor is the noise present in the data. For extended radio sources, this error is characterised by the rms in the image multiplied with the square root of the ratio of the solid angle (of the extended source) to that of the synthesised beam, which is the number of beams across the extended source. In addition to this, we also included uncertainty due to point source subtraction from {\it uv} data \citep{2019SSRv..215...16V}. These three uncertainties factors are unrelated; therefore they are added in quadrature to estimate the absolute error on the flux densities of the extended sources, as shown below:
\begin{equation}
\Delta{S} = [(\sigma_{\mathrm{amp}}S)^2 + (\sigma_{\mathrm{rms}}\sqrt{n_{\mathrm{beams,ext}}})^2) + \sigma_{sub}^{2}]^{1/2},
\label{cal_err}
\end{equation}
  \par where $S$ is the flux density, $\sigma_{\mathrm{amp}}$ is the flux calibration uncertainty (10\%), $\sigma_{\mathrm{rms}}$ is the image rms noise, and $n_{\mathrm{beams,ext}}$ is the number of beams in the full extent of the extended source, $\sigma_{sub}^{2}$ = $\sum_{i} (\sigma_{\mathrm{rms}}\sqrt{n_{\mathrm{beams,pt}}})^2$, where $n_{\mathrm{beams,pt}}$ is the number of beams in the subtracted point sources.

The GMRT (325 MHz) and MeerKAT flux densities are on the PB10 and \cite{ReyATNF.352..508R} scales, respectively. The agreement between these two scales is within 2\% (private communication with MeerKAT commissioning team).  
For spectral index measurements (\S \ref{spxplot}), we took measured flux densities at 241 and 610 MHz from the V09. The authors used the \cite{1977A&A....61...99B} standard; hence we scaled the 325 MHz flux densities by the ratio between the \cite{1977A&A....61...99B} to PB10 standards.  


\subsection{Direction-dependent calibration}
\label{ddcal}
\par Radio interferometric wide-band and wide-field of view observations are often dynamic range limited. A number of instrumental and observational errors affect the quality  of such radio observations. These errors typically manifest themselves as artefacts associated with bright and strong sources in the observed field, particularly off-axis sources. These artefacts can extend throughout the map (being, in a sense, modulated by the extensive sidelobes of the radio interferometer point spread function (PSF)) and therefore hamper the detection of faint and diffuse emission and imaging of extended radio source(s) elsewhere in the map (including at the centre). There are several techniques available to minimise this problem by employing some form of direction-dependent calibration \citep[e.g.][]{2019RNAAS...3..110W,2011MNRAS.414.1656K,2011A&A...527A.107S, 2009A&A...501.1185I}. One of the first such methods proposed was ``source peeling'' which deals with strong sources by solving for the gain towards them one by one, followed by the source subtraction from the visibility ({\it uv}) data. A generalization of this is the ``differential gains'' technique \citep{2011A&A...527A.107S,2011A&A...527A.108S}, which can be seen as a form of simultaneous peeling, calibrating and correcting towards multiple bright sources simultaneously.  In this work, we employed the CubiCal software \citep{2018MNRAS.478.2399K} in conjunction with the DDFacet imaging software \citep{2018A&A...611A..87T} to peel bright and strong sources. CubiCal solves the following Radio Interferometer Measurement Equation (RIME)       
\begin{equation} 
D_{p,q} =  G_{p}(\sum_{i=1}^{n}dE^{(i)}_{p}P^{(i)}_{p}S_{p,q}^{(i)}P_{q}^{^{(i)}H}dE_{q}^{^{(i)}H}+S_{p,q}^{\mathrm{(di)}})G_{q}^{H}
\end{equation}
where $G$ terms are the direction-independent errors affecting the entire field of view. In our case, we constrain these to phase-only solutions (the equivalent of traditional phase selfcal). The $dE$ terms represent direction-dependent effects associated with $i=1...n$ bright sources. These use full complex 2x2 Jones matrix solutions \citep{2015MNRAS.449.2668S}. The $P$ terms implemented antenna-specific primary beam models (time and frequency varying). The $S^{(i)}$ terms denote \emph{intrinsic} (and direction-dependent) sky models for the individual sources being peeled, while $S^{\mathrm{(di)}}$ represents the ``direction-independent'' model for the rest of the emission in the field, derived from clean model components. 

The sky model is generated using the DDFacet imaging software, and source-specific models (direction-dependent) are predicted using on-the-fly targeted faceting from subsets of the sky model. CubiCal can solve for both of these ($G$ and $dE$) errors simultaneously, while taking into account most of the sky flux. This limits the flux suppression of underlying radio sources (compact and extended sources). 

Firstly, self-calibrated visibilities (i.e. with a $G$ solution only) are used to generate an initial sky model with DDFacet. We then designate regions (using the SAOImage ds9 viewer) corresponding to the individual bright sources that require peeling. CubiCal then uses these regions to split the sky model predict into $S^{(i)}$ and $S^{\mathrm{(di)}}$ components, and performs simultaneous $G$ and $dE$ solutions as per the RIME above. The bright sources (using their best-fitting $dE$ solutions) are subtracted from the visibilities, the residuals are corrected for $G$, and then re-imaged again using DDFacet. This procedure results in images with the bright sources (and associated artefacts) mostly removed, with improved image quality.
\subsection{Point source subtraction}
\label{ptsub}
\par In radio interferometry observations generally, unresolved and discrete point sources are blended with extended diffuse radio emission. Sometimes, the brightness of diffuse emission is obscured due to the presence of very bright point sources, and hence it is not easy to image the extended sources. In order to estimate the flux densities of extended sources, we need to subtract these point sources blended with the diffuse emission. Removing point sources is also an important task for generating a high-resolution spectral index map. There are various techniques available for excising point sources. A standard method is to use the box cleaning option in various clean algorithms. These boxes can be used for (affecting) strong point sources while cleaning and generating the model, which will later be subtracted from calibrated data. The other widely used method is to image the data with uniform weighting (and exclude the inner interferometry baselines that correspond to the angular size of the extended source that needs to be image) to generate the high-resolution map. This high-resolution data should contain only point sources and resolve out any extended emission. This will create models (of only point sources) which can then be subtracted from the calibrated data before re-imaging the residual visibility at lower resolution. Both methods, however, depend on the different angular scales of the underlying radio sources and are performed iteratively. They are, therefore, time-consuming when the data volume is large. Hence, it is necessary to develop a method that can subtract the point sources efficiently in a standard imaging procedure without more iterations. To do this, we used the Crystalball software\footnote[3]{https://github.com/paoloserra/crystalball} which can be employed in a Stimela script (or pipeline) to perform the point source subtraction. In this procedure, a user has to first make a region file that lists the coordinates of point sources which they want to remove. This region file can be manually generated via the SAOImage ds9 viewer. The user then runs WSClean with the \texttt{-save-source-list} option (in the final self-calibration imaging). This will generate a clean components file\footnote[4]{https://sourceforge.net/p/wsclean/wiki/ComponentList/}. Crystalball takes these two files, point source regions and clean component list, and generates another column in a given Measurement Set. This can be subtracted from the visibility and the residuals re-imaged. In our case, we made a source list from a high-resolution map of the ZwCl2341.1+0000 (uniform weighting and excluding inner MeerKAT baselines of 1 km resulting in 5$''$ beam size). In this high-resolution map, there are 1800 point sources that have been detected in the PyBDSF \citep{2015ascl.soft02007M} generated catalogue. We selected only compact sources i) detected in the high-resolution image (uniform weighting), ii) blended in double radio relics and nearby strong sources (Figure \ref{dd_labled_marked}) and iii) situated in between relics region. This technique of point source subtraction from visibilities is very efficient in the removal of many point sources simultaneously. The advantage of this method is that the Crystalball software does a direct Fourier transform (DFT) on the clean component list generated by the WSClean. This is more accurate than using model images because it can predict a smooth spectral polynomial at native channel resolution. The other advantage of this method is that all the required software are part of the Stimela container; hence one can implement all mentioned tasks in a pipeline. 

\section{Results}
\par
\begin{center}
\begin{figure*}
    \centering
    \begin{subfigure}[t]{0.45\textwidth}
        \includegraphics[width=1\textwidth]{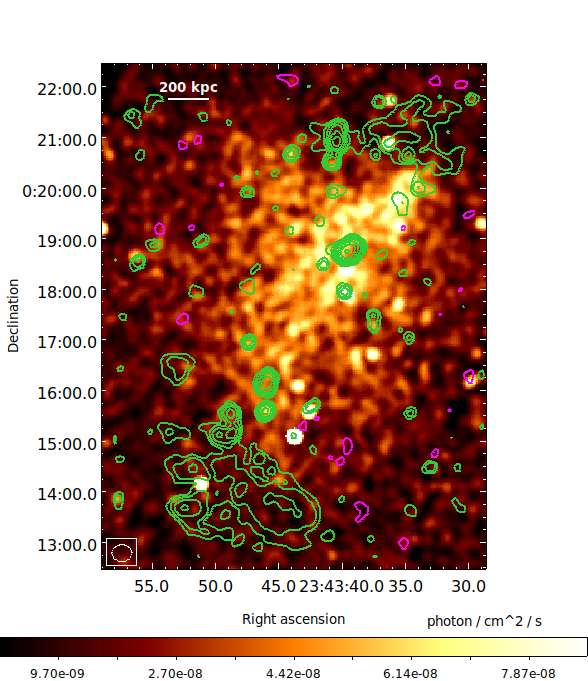}
        \caption{}
        \label{rfidtest_xaxis1}
    \end{subfigure}
    \begin{subfigure}[t]{0.45\textwidth}
        \includegraphics[width=1\textwidth]{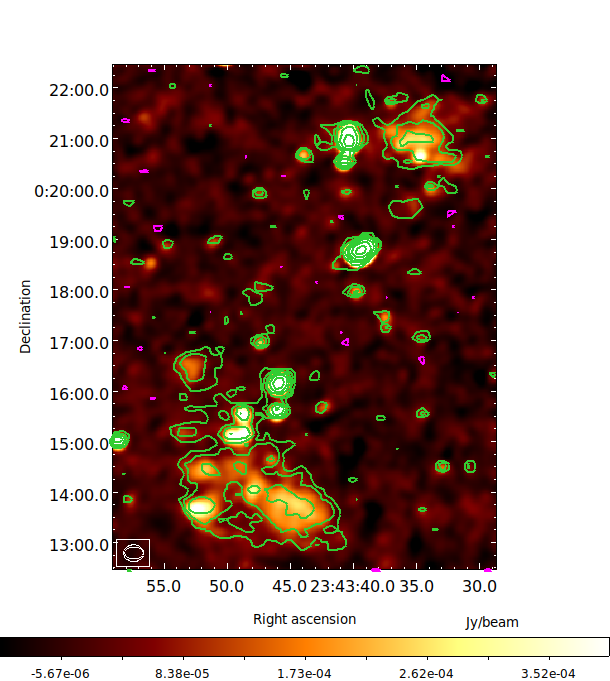}
        \caption{}
        \label{rfidtest_yaxis2}
        \end{subfigure}
    \caption{(a) ZwCl 2341+0000 {\it Chandra} smoothed X-ray color image. The green contours show the MeerKAT observation. (b) GMRT 325 MHz radio contours on MeerKAT radio color image. In both images, the contour levels are [1,2,4,8,16,32,64,128] $\times$ 3$\sigma$. Negative contours are shown by the magenta color. For the MeerKAT radio image, the beam size is 12$''$ $\times$ 10$''$ and 1$\sigma$ = 20 $\mu$Jy beam$^{-1}$. In the GMRT radio image, the beam size is 11$''$ $\times$ 7$''$ and 1$\sigma$ = 30 $\mu$Jy beam$^{-1}$.}
    \label{Xray_radio_img}        
\end{figure*}
\end{center}

\begin{center}
\begin{figure*}
\centering
\includegraphics[width=1\textwidth]{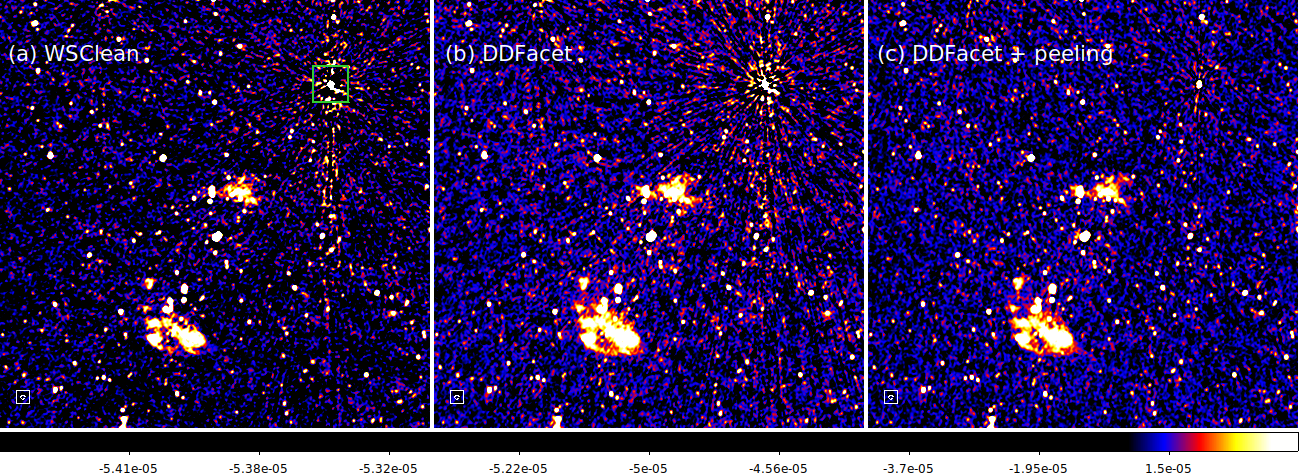}
\caption{Results of direction-dependant calibration. The left image (only shown central part of ZwCl 2341.1+0000) is the self-calibrated direction-independent image generated from the CARACal. The middle image is the self-calibrated image generated using the DDFacet. The right image is the direction-dependent image (after calibrating towards five directions using the CubiCal). In the left image, we marked the bright and strong radio source in the green box, which affects the region of relics. All images have the same resolution of 8$''$ $\times$ 6$''$ and colour scale. We calculated the local rms, minimum pixel and sum of total negative component values inside the 8$'$ box size around the marked strong source before and after direction-dependent calibration. These values are listed in Table \ref{DD_stat}.}
\label{dd_calibration}        
\end{figure*}
\end{center}

\begin{center}
\begin{figure*}
    \centering
    \begin{subfigure}[t]{0.55\textwidth}
        \includegraphics[width=1\textwidth]{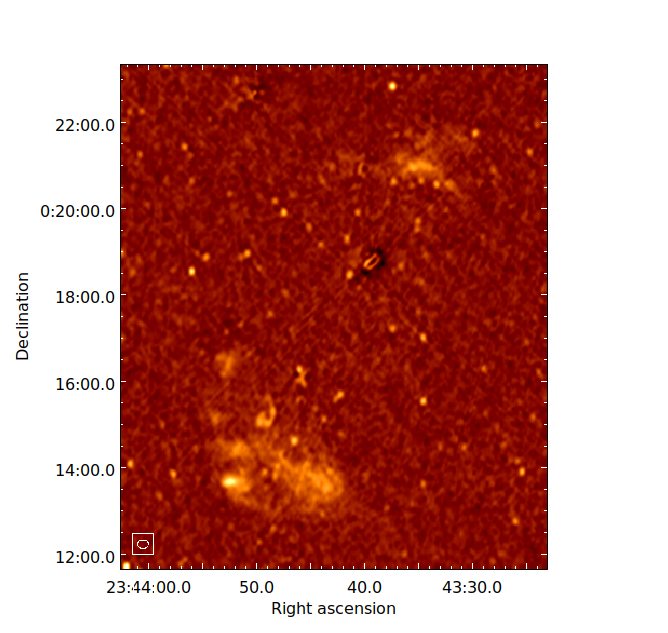}
        \caption{}
        \label{rfidtest_xaxis1}
    \end{subfigure}
    \hspace{-2cm}
    \begin{subfigure}[t]{0.55\textwidth}
        \includegraphics[width=1\textwidth]{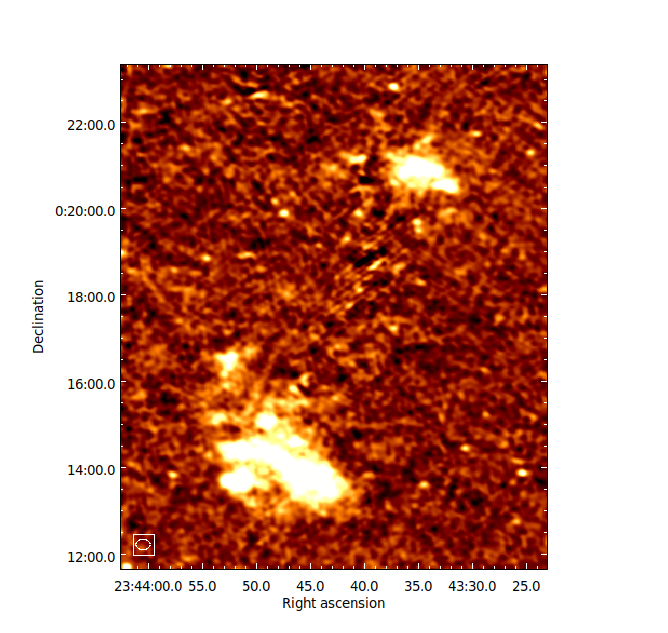}
        \caption{}
        \label{rfidtest_yaxis2}
        \end{subfigure}
    \caption{Discrete point sources around radio relics have been subtracted for (a) MeerKAT (direction-dependent image) and (b) GMRT (direction-independent image). For the MeerKAT radio image, the beam size is 8$''$ $\times$ 6$''$  and in the GMRT radio image, the beam size is 11$''$ $\times$ 7$''$.}
    \label{radio_pt_img}        
\end{figure*}
\end{center}

 We have shown our MeerKAT results from Figure \ref{Xray_radio_img} to Figure \ref{radio_pt_img} as well as from Figure \ref{dd_labled_marked} to Figure \ref{A2631_dde_compt_src}.  In this section, our main focus is to study the diffuse and extended radio sources after direction-dependent calibration. There are double radio relics that have already been discovered in the ZwCl 2341.1+0000 \citep{2002NewA....7..249B}. We show the MeerKAT and GMRT direction-independent images of double radio relics in Figure \ref{Xray_radio_img}. To compare and match the resolutions, we tapered the MeerKAT image (\texttt{robust = 0}) with 10$''$ to match the beam size of the GMRT image.  As previously shown in the GMRT and VLA observations (G10,V09), there are two double radio relics; one in the north-west and another in the south-east directions clearly visible in the MeerKAT L-band and 325 MHz GMRT data. Similar to the previous observations, these two relics are situated perpendicular to the merger axis (north-south). Furthermore, they are located outside the hot and dense X-ray intra-cluster medium (ICM) region at the periphery of the merger. Both relics have a regular symmetric structure, slightly deviating from the usual elongated arc-like structure attributed to relics. We also noticed that the southern relic is larger (496 $\times$ 1241 kpc$^{2}$) in size and more elongated compared to the northern relic (270 $\times$ 426 kpc$^{2}$). The size of the southern relic is similar to the 610 MHz observation as measured by V09, but the linear size of the northern relic in the MeerKAT observation is larger than the 610 MHz measurements.

In the MeerKAT data, there is a strong and unresolved point source (RA:23:43:16.569, Dec:+0:26:20.392) located at a distance of $\sim$ 7$'$ ($\sim$ 1.7 Mpc) from the north relic (NR), $\sim$ 15$'$ ($\sim$ 3.7 Mpc) from the south relic (SR) and $\sim$ 9$'$ from the phase-centre (pointing centre) of the ZwCl 2341.1+0000 field,  marked in the green box region in Figure \ref{dd_calibration} (left). The total flux density of this radio source is $\sim$ 87 mJy, and it causes artefacts and hence poor image quality around the central part of the cluster where two relics are situated. In Figure \ref{dd_calibration} (left) shows the CARACal image (final self-calibrated) generated with WSClean. As seen in this Figure, clearly the artefacts are affecting both north and south radio relics. Then in addition to this strong source, we also chose four different strong point sources or directions to apply the direction-dependent calibration. We used DDFacet to generate the facet based initial sky model required to calibrate the data in different directions. Figure \ref{dd_calibration} (middle) shows the DDFacet generated image which is still uncalibrated for distinct directions. Finally, Figure \ref{dd_calibration} (right) shows the direction-dependent calibrated image. We have seen clear improvement around the relic region before and after the direction-dependent calibration. We found a 25$\%$ rms noise decrease in the region between two relics after the direction-dependent calibration. All these maps are made at \texttt{robust = 0} value. This results in a FWHM of 8$''$ $\times$ 6$''$. Figure \ref{zwcl_dd_calibration} shows the full field of the ZwCl 2341.1+0000 before (left image) and after (right image) the direction-dependent calibration. In Figure \ref{zwcl_dde_compt_src}, we have shown four different point sources towards calibration applied and peeled the sources in order to improve the image quality. Similar to this, we also applied direction-dependent calibration to the A2631 field and results are shown in Figure \ref{A2631_dd_calibration} and Figure \ref{A2631_dde_compt_src}. In both of these fields, for every peeled source, we estimated local rms, minimum clean component value and sum of the total negative components around 8$'$ region. We have given these values in Table \ref{DD_stat}. 

 In the ZwCl2341.1+0000 radio map, we noticed discrete and blended point sources as shown in Figure \ref{dd_labled_marked}. Some of them are also reported in V09 which we labelled in Figure \ref{dd_labled_marked}(left). Figure \ref{dd_labled_marked}(right) shows the image of only discrete point sources, generated using the Crystalball (as mentioned in \S \ref{ptsub}). We subtracted these point sources from the MeerKAT (direction-dependent calibrated) and GMRT visibilities using our point source subtraction technique. We listed the properties of subtracted point sources in Table \ref{pt_list}. We have shown results in Figure \ref{radio_pt_img}(a) and (b) for MeerKAT and GMRT, respectively. We calculated flux densities of both radio relics and reported these values in Table \ref{relic_prop}. We also compared flux densities of compact radio sources before and after direction-dependent calibration and showed  the result in Figure ~\ref{dd_calibration_compare}. We found an agreement within 2$\%$; if we include only bright point sources (of $>$ 1 mJy), then the agreement is within 0.5\%. 
 \par {In our forthcoming papers, we will give (1) a detailed analysis of these relics and the MeerKAT sub-band spectral index map (Kincaid et al. in prep.) and (2) radio source identifications and catalogue, spectral indices, and overall radio galaxy populations and their properties (for clusters - ZwCl 2341.1+0000 and A2631) in the core region of the Saraswati Supercluster (Parekh et al. in prep.)} 

\begin{table}
\caption{Flux density measurements of diffuse radio sources.}
\label{relic_prop}
\begin{tabular}{ccccccccc}
\hline
\hline
Frequency (MHz)  & 343 &  1283 \\
\hline
Synthesized beam ($''$ $\times$ $''$,$^{\circ}$)  & 11 $\times$ 8, 79  & 8 $\times$ 6, -3 \\
rms ($\mu$Jy beam$^{-1}$)               & 30   & 14  \\
\hline
North relic (mJy) &    18.0 $\pm$ 1.85        & 5.0 $\pm$ 0.51 \\
South relic (mJy) &    64.0 $\pm$ 6.41         & 17.0 $\pm$ 1.72\\
\hline
Synthesized beam ($''$ $\times$ $''$,$^{\circ}$)  &   & 50 $\times$ 50, 0 \\
rms ($\mu$Jy beam$^{-1}$)               &    & 77  \\
\hline
Halo (mJy)        &          -               & 3.3  $\pm$ 0.50 \\
\hline

\end{tabular}
\end{table}

\vspace{-1.5cm}

\section{Is there diffuse halo present in ZwCl2341.1+0000?}
\par
\begin{center}
\begin{figure}
\centering
\includegraphics[width=0.5\textwidth]{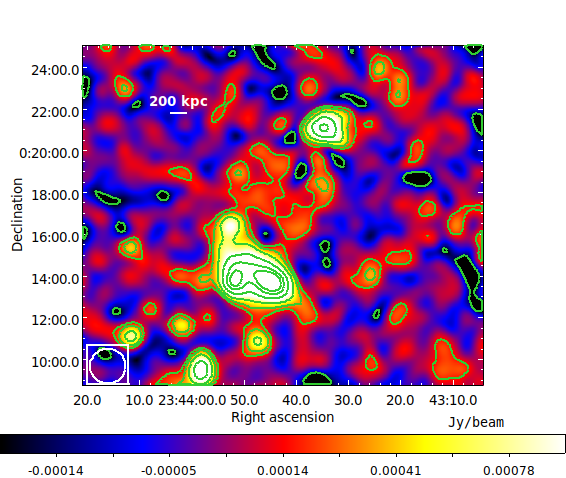}
\caption{MeerKAT image of ZwCl 2341.1+0000 with 50$''$ beam size and rms sensitivity $1\sigma = 77 \mu$Jy beam$^{-1}$ (after direction-dependent calibration). The Contour levels are $[1,3,5,7,9,11] \times 2 \sigma$.  The first contour is drawn at 2$\sigma$.}
\label{mk_halo_img}        
\end{figure}
\end{center}

 In the high-resolution (\texttt{robust = 0}) MeerKAT observation of ZwCl2341.1+0000, we could not detect any large radio emission between two radio relics.  We applied {\it{uv}}-tapering to the MeerKAT data to match the resolution with the NVSS where \cite{2002NewA....7..249B} have been known to show the possibility of an extended radio bridge between relics. We generated images with different tapering using the direction-dependent calibrated- and point source subtracted data. At 50$''$ resolution, as shown in Figure ~\ref{mk_halo_img}, we detected a radio source between double radio relics.  This suggests that the specific radio emission is resolved out in the high-resolution images. Based on its position, it could be a candidate radio halo, however, the nature of this radio source is not clear. The radio emission visible in the MeerKAT data is irregular and complex. This radio emission is patchy and faint (surface brightness of $\sim$ 0.13 mJy beam$^{-1}$) and extended around the central region of ZwCl2341.1+0000, connected with both radio relics.  The detection level of this diffuse emission is only 2$\sigma$ in the present data. The rms of the image is 77 $\mu$Jy beam$^{-1}$ which is a factor of two better than previous detections in the VLA (L-band, D-configuration) as reported in G10, and five times better than the NVSS. Their beam sizes, however, were $\sim$ 80$''$ and 45$''$, respectively. We have not detected a halo emission in the 80$''$ beam size of the MeerKAT data (of which rms $\sim$ 0.2 mJy beam$^{-1}$).  We found the size of the possible candidate halo (at 2$\sigma$ level) is $\sim$ 221$''$ $\times$ 286$''$ (914 $\times$ 1183 kpc$^{2}$) which is comparable to typical radio halo sizes. The flux density within 2$\sigma$ level is reported in Table \ref{relic_prop}. It is difficult to study its morphology and other radio properties in the current data. More data is needed to confirm this detection.

\section{Spectral index plot of relics and P1.4 GHz values}
\label{spxplot}
\par
\begin{center}
\begin{figure}
\centering
\includegraphics[width=0.4\textwidth]{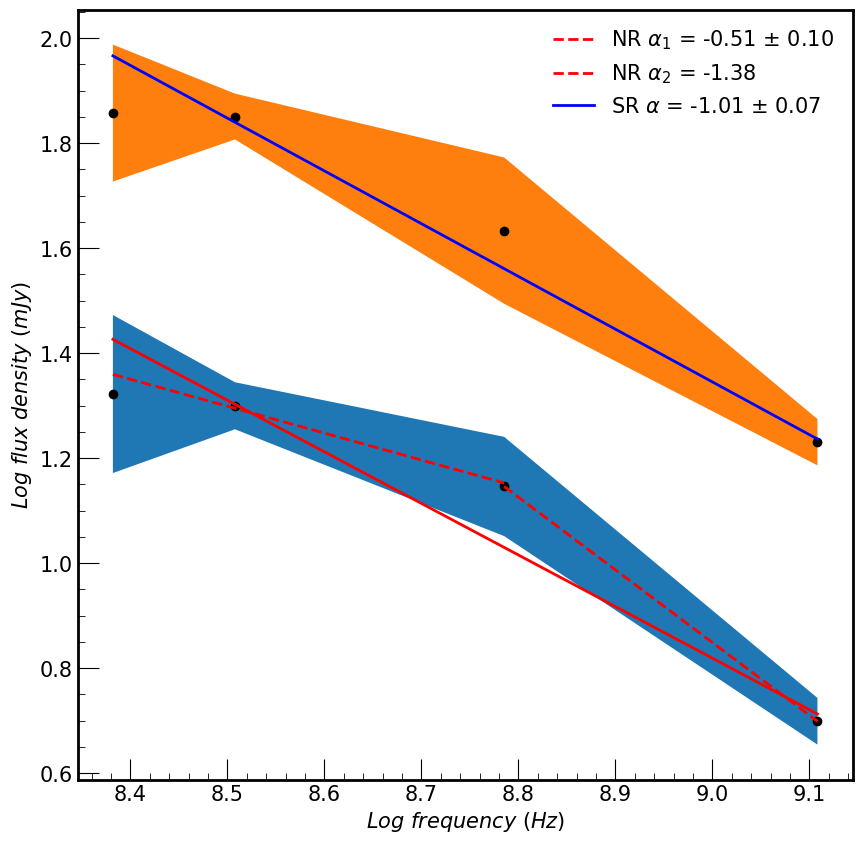}
\caption{1D spectral index plot for north relic (NR) and south relic (SR). {The red solid line shows the best-fit line between 241 and 1283 MHz for the NR}.}
\label{1d_spx}        
\end{figure}
\end{center}

 We have shown a spectral index plot of both radio relics in Figure ~\ref{1d_spx}. We derived the spectra between 241 to 1283 MHz. We took 241 and 610 MHz measurements from V09 and 323 and 1283 MHz from this work. For the SR, we combined flux densities of regions RS-1, RS-2, and RS-3 as labelled by V09. We matched the different calibration scales (\S \ref{cal_scale}) in order to estimate flux densities. {The minimum separation between antenna pairs of MeerKAT is 29 m, hence MeerKAT is insensitive to image a structure larger than $\sim$ 27$'$ at 1283 MHz. Similarly, GMRT is sensitive only up to $\sim$ 32$'$and 17$'$ at 325 and 610 MHz bands, respectively. This should be considered while comparing flux densities of extended sources using different arrays. In our case, the sizes of both relics are less than these largest detectable structures of different data sets.} We found a straight power-law fit for the SR of spectral index value $\alpha_{\mathrm{south}}$ = -1.01 $\pm$ 0.08. For the NR, we found a break at 610 MHz, and the single power-law model ({red solid line in Figure~\ref{1d_spx}, of $\alpha$ = -0.98 $\pm$ 0.10}) does not fit the data. We found a flat spectrum between 241 to 610 MHz, which became steep thereafter. We measured two slopes, $\alpha_{\mathrm{1,north}}$ = -0.51 $\pm$ 0.10 and $\alpha_{\mathrm{2,north}}$ = -1.38. The value of $\alpha_{\mathrm{1,north}}$ is similar to the value given by V09 ($\alpha_{\mathrm{north}}$ = -0.49 $\pm$ 0.18).

\par For double radio relics and the candidate halo, we estimated its K-corrected (rest-frame) radio power ($P_{1.4\mathrm{GHz}}$) using the following: 
\begin{equation}
P_{1.4\mathrm{GHz}} = \frac{4 \pi D_{\mathrm{L}}^{2}S_{v}}{(1+z)^{1+\alpha}},
\end{equation}
\par where $D_{\mathrm{L}}$ is the source luminosity distance at $z$ = 0.27,  and $S_{v}$ is the source flux density. For the north radio relic, $\alpha$ $\sim$ -1.3 (between 610 and 1283 MHz) gives radio power at 1.4 GHz to be $P_{1.4\mathrm{GHz}}$ = (1.08 $\pm$ 0.11) $\times$ 10$^{24}$ W Hz$^{-1}$. For the south radio relic, $\alpha$ $\sim$ -1.0 gives radio power to be $P_{1.4\mathrm{GHz}}$ = (3.53 $\pm$ 0.35) $\times$ 10$^{24}$ W Hz$^{-1}$. For radio halo, $P_{1.4\mathrm{GHz}}$ = (7.17 $\pm$ 1.08) $\times$ 10$^{23}$ W Hz$^{-1}$. We used $\alpha$ = -1.3 to scale the halo flux density to 1.4 GHz.

\section{Discussion and conclusion}
\par Using the MeerKAT telescope, we observed the centrally surrounded two massive galaxy clusters, A2631 and ZwCl 2341.1+0000, of the {\it Saraswati} supercluster. In this work, we aimed to test the direction-dependent calibrations on wide-field of view and broadband MeerKAT data and show the improvement in diffuse radio source imaging. Furthermore, we also studied particular of the ZwCl2341.1+0000 merging galaxy cluster. The cluster is elongated and displays the disturbances in X-ray emission. For more details about the dynamics and merging geometry of the cluster, we refer to \cite{2014MNRAS.443.2463O}. Previously, in low frequency and high-resolution GMRT data, two radio relics have been detected (V09). NVSS and low-resolution VLA data also suggested the presence of relics and extended diffuse radio source in the cosmic web environments in overlap with a chain of galaxies.

\subsection{Direction-dependent effects}
\par The peeling techniques we have employed to improve the image qualities of our MeerKAT data are discussed in \S \ref{ddcal}. In our direction-dependent analysis, we found that within 35$'$ from the phase centre,  strong, poorly deconvolved sources can be peeled quite well, with considerable attenuation of the associated artefacts. 

Figure \ref{zwcl_dde_compt_src} shows the peeled sources, ordered in increasing distance from centre. Our first peeled source (shown in Figure \ref{dd_calibration}) is 8$'$ off-axis, and was dramatically improved by the direction-dependent calibration. The other four sources (Figure \ref{zwcl_dde_compt_src}(a), (b), (c) and (d)) are at distances of 12$'$, 26$'$, 31$'$ and 38$'$, respectively. As one can see, sources within $<$ 35$'$ can be improved. For the other field of A2631, the three selected strong sources (Figure \ref{A2631_dde_compt_src}(a), (b) and (c)) are at distances of 23$'$, 30$'$ and 51$'$, respectively, from the phase-centre. As one can see, the level of artefacts around the first two sources are improved after the direction-dependent calibrations, whilst residual artefacts are still visible (albeit at a low level) for the third source. As a quantitative measure, we calculate the local rms, minimum pixel and sum of total negative pixel values in an $8'\times 8'$ box centred on each peeled source (Table \ref{DD_stat}). The rationale for looking at negative pixels is that imaging artefacts due to calibration errors manifest themselves in a positive-negative pattern (fundamentally, because PSF sidelobes are positive-negative), while physical emission is all positive. The distribution of the negative pixels is therefore indicative of the artefact level. As we can see in Table \ref{DD_stat} (sources are given in order of increasing distance from centre), there are clear decrements in the (absolute) sum of negative values after direction-dependent calibration. Local rms is also improved after the direction-dependent calibration, and (absolute) minimum pixel values are also reduced after the direction-dependent calibration, although this does not seem to be a significant indicator. 

\par The success of the peeling process will also depend on source location in the given image as well as source structure. Since MeerKAT provides wide-bandwidth observations, it is important to account for the variation of the primary beam across the broad frequency range. The size of the primary beam and its half-power points vary radially with the frequencies within the MeerKAT band. Furthermore, if antenna beam patterns are not modelled well across the wide bandwidth, then accuracy in the estimate of the flux densities of off-axis sources (away from the phase-centre) decrease \citep{10.1093/mnras/stz702}.  Hence, the calibration process computes incorrect gains for these sources. This makes it difficult to improve residual artefacts around strong sources situated at the edges of image. With the advancement in MeerKAT future beam modeling, we can hope to further alleviate the artefacts associated with such edge sources.

\subsection{Diffuse radio sources in ZwCl2341.1+0000}
\par The candidate radio halo (or bridge) in ZwCl2341.1+0000 was first detected in the NVSS survey which provides a beam size of $\sim$ 45$''$ and was not detected in high-resolution GMRT observation, even at low frequencies. In low-resolution VLA L-band observation of ZwCl2341.1+0000 (G10), it was not possible to separate three diffuse components of the north and south radio relics and candidate radio halo lying between them. Our MeerKAT L-band observation provides the highest resolution and sensitivity compared to the previous 1.4 GHz observations.  We could not detect the candidate radio halo in a high-resolution (8$''$ $\times$ 6$''$) MeerKAT image. However, we do see a tentative detection of it at a 2 sigma significance level after tapering the MeerKAT image to a 50$''$ beam size. The candidate halo emission is resolved out in high-resolution data and could be detected with compact array configurations only where extended radio emissions have higher surface brightness. The radio halo emission in the MeerKAT observation is very irregular and patchy. The rms statistics and detection level are better in the direction-dependent tapered image than the direction-independent image. The tapering of the {\it uv} data increases the rms noise; hence we could not taper to larger beam sizes to improve the surface brightness of the halo emission. Based on the cluster X-ray mass and halo size scaling relationship \citep{2013ApJ...777..141C,2007MNRAS.378.1565C}, the expected linear size of the halo is $\sim$ 756 kpc, however, the largest linear size of the halo that we could measure was $\sim$ 1183 kpc. This is comparable to the size estimated by G10. The flux density measurement of the halo in MeerKAT observation is $\sim$ 3 times lower than that reported by G10. This could be due to the difference in beam sizes and array configurations of MeerKAT and VLA. The expected power of the radio halo based on the scaling relation \citep{2013ApJ...777..141C} is $\sim$ 8.5 $\times$ 10$^{23}$ W Hz$^{-1}$, which is close to our measurement (7.17 $\pm$ 1.08) $\times$ 10$^{23}$ W Hz$^{-1}$, in MeerKAT data. {It has been shown that the radio power of giant radio halos is correlated with the X-ray masses of their host clusters \citep{osti_22518716,2000ApJ...544..686L}. Based on this, the radio power of the candidate halo of ZwCl 2341.1+0000 is consistent with the $P_{1.4GHz}$ vs Mass relation of giant halo clusters}. However, the nature of this candidate radio halo emission is still unclear and it could also be related to the cosmic web network of the {\it Saraswati} supercluster. Its formation mechanism could also be related to the LSS formation instead of cluster merger. More data is required to characterise the properties of this puzzling candidate radio halo.  
\par G10 have reported the total flux density of all three components as $\sim$ 28.5 mJy. In our observation, we found total flux densities of the double radio relics and halo are = 25.3 $\pm$ 1.86 mJy, which is comparable. G10 measured the total linear size of all three diffuse components as $\sim$ 2.2 Mpc, while we estimated the total linear size as $\sim$ 2.4 Mpc. We also noticed that the NR (in MeerKAT observation) has a larger linear size than compared to the 610 MHz data (V09), but we also note that these  observations have different sensitivities and background rms values. This could suggest that our sensitive MeerKAT observations have detected more diffuse emissions in the northern radio relic (towards the north direction) than previous observations. { Based on radio relics cluster $P_{1.4GHz}$ vs Mass relationship \citep{2017MNRAS.472..940K}, the radio power of both NR and SR are consistent with this relation}. Furthermore, G10 reported the spectral index value between 610 and 1400 MHz for north and south relics as $\sim$ -1.2. We fitted a straight line with spectral index, $\alpha_{1283\mathrm{}}^{610\mathrm{}}$ $\sim$  -1.2 for the SR, which is comparable with G10. The spectral index value for SR, $\alpha_{1283\mathrm{}}^{241\mathrm{}}$ = -1.01 $\pm$ 0.08. {Based on radio relic spectral indices compilation \citep{2021MNRAS.506..396W}, we found an average spectral index value of known radio relics is $\sim$ -1.2. The spectral index value of the SR is comparable to this average value}. The NR is faint compared to the SR, and its spectrum appears complex (Figure \ref{1d_spx}). There is a spectral break at 610 MHz with a steeper spectrum ($\alpha_{1400\mathrm{}}^{610\mathrm{}}$ $\sim$ -1.38) towards the high frequency and a flattening towards lower frequencies ($\alpha_{610\mathrm{}}^{241\mathrm{}}$ = -0.51 $\pm$ 0.10). {This spectral break, if confirmed with further sensitive low frequency data, then is very unusual and inconsistent with the diffusive shock acceleration (DSA) mechanism \citep{1983RPPh...46..973D}, which advocates the single power-law distributions for relic spectra. In the last decade, several radio relics have been observed over $>$ 1 GHz frequency bands \citep{2015A&A...575A..45T,2014MNRAS.441L..41S,2014ApJ...794...24O,2013A&A...555A.110S}. It was found that the A2256 also shows uncommon spectral break at $>$ 1.4 GHz \citep{2015A&A...575A..45T}. This indicates the halting of injection of electrons in the relic system, which eventually leads to cutoff at high frequencies of the integrated spectrum. The alternative explanation could be for this spectral break is clusters merging process is happening recently (young merger) where relics have just formed, and the electrons have not yet had time to lose their energy to steepen the spectrum. We will study this spectral break in more detail in our forthcoming paper (Kincaid et al. in prep.).} We emphasize that all different instruments (MeerKAT, VLA and GMRT) have different {\it uv} coverage, antenna configurations, observation conditions and sensitivities; hence our spectral indices comparisons with other observations are affected by these constraints.      

\par In this work, we have shown the direction-dependent calibration technique for galaxy cluster fields. It both improves the image quality and allow us to estimate the flux densities of diffuse radio sources after eliminating artefacts caused by strong sources. This method is important in the wide-bandwidth and wide-field of view observations of superclusters or galaxy clusters in order to detect faint and diffuse radio emissions in the presence of strong radio source(s). This  technique can be implemented in a pipeline and can be applied to other data. In our next paper, we will study the merger shock properties in the supercluster environment and broadband spectral index studies of diffuse radio sources.  
\\
\\
\noindent{\it Acknowledgements.} The financial assistance of the South African Radio Astronomy Observatory (SARAO) towards this research is hereby acknowledged. The MeerKAT telescope is operated by the South African Radio Astronomy Observatory, which is a facility of the National Research Foundation, an agency of the Department of Science and Innovation. This work is based upon research supported by the South African Research Chairs Initiative of the Department of Science and Technology and National Research Foundation. V. Parekh acknowledge Hiroki Akamatsu, Bill Cotton and Annalisa Bonafede for fruitful discussion. (Part of) the data published here have been reduced using the CARACal pipeline, partially supported by ERC Starting grant number 679629 ``FORNAX", MAECI Grant Number ZA18GR02, DST-NRF Grant Number 113121 as part of the ISARP Joint Research Scheme, and BMBF project 05A17PC2 for MeerKAT. Information about CARACal can be obtained online under the URL: https://caracal.readthedocs.io.
RK acknowledges the support of the Department of Atomic Energy, Government of India  
 under project no. 12-R\&D-TFR-5.02-0700. SS acknowledges the support of the European Regional Development Fund and the Mobilitas Pluss postdoctoral research grant MOBJD660.
\\
\noindent{\it Data Availability.} The data underlying this article are subject to an embargo. Once the embargo expires the data will be available [https://archive.sarao.ac.za/; https://www.sarao.ac.za/wp-content/uploads/2019/12/MeerKAT-Telescope-and-Data-Access-Guidelines.pdf].

\bibliography{references,ref_shishir}

\newpage
\appendix
\counterwithin{figure}{section}
\counterwithin{table}{section}

\section{Appendix}
\newpage
\begin{figure*}
\includegraphics[width=0.55\textwidth]{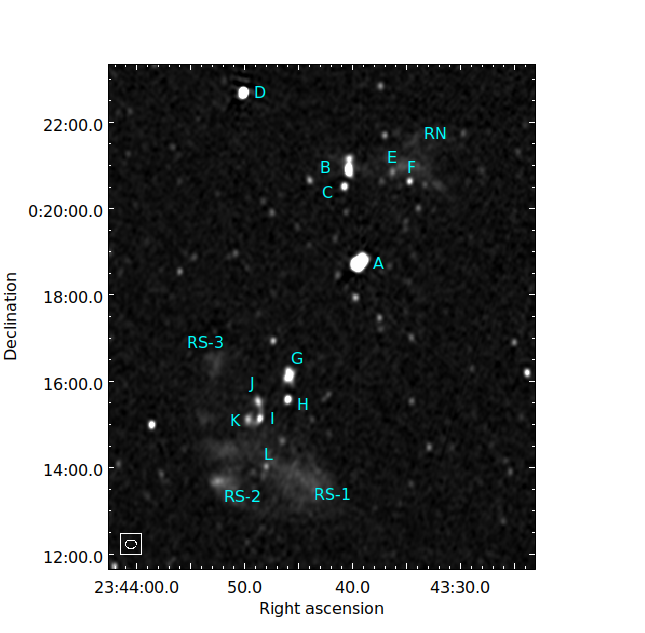}
\hspace{-2cm}
\includegraphics[width=0.55\textwidth]{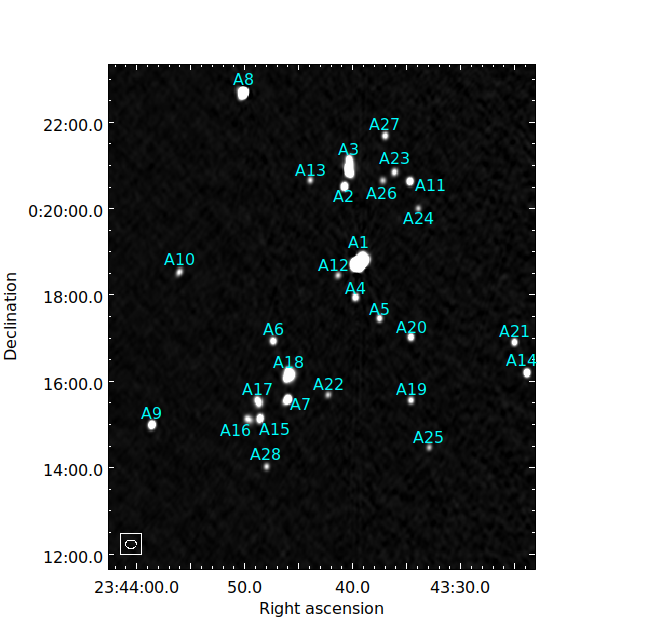}
\caption{(left) MeerKAT direction-dependent calibration image of ZwCl2341.1+0000 (\texttt{robust = 0}). We marked sources with alphabets as labelled by V09. The beam size is 8$''$ $\times$ 6$''$, pa = -3$^{\circ}$ and (global) rms is 14 $\mu$Jy beam$^{-1}$. (right) Same as the left image, but only showing the compact point sources (generated with the Crystalball) which were subtracted from the visibility. We marked each source from A1 to A28.}
\label{dd_labled_marked}        
\end{figure*}

\begin{center}
\begin{figure*}
\includegraphics[width=1\textwidth]{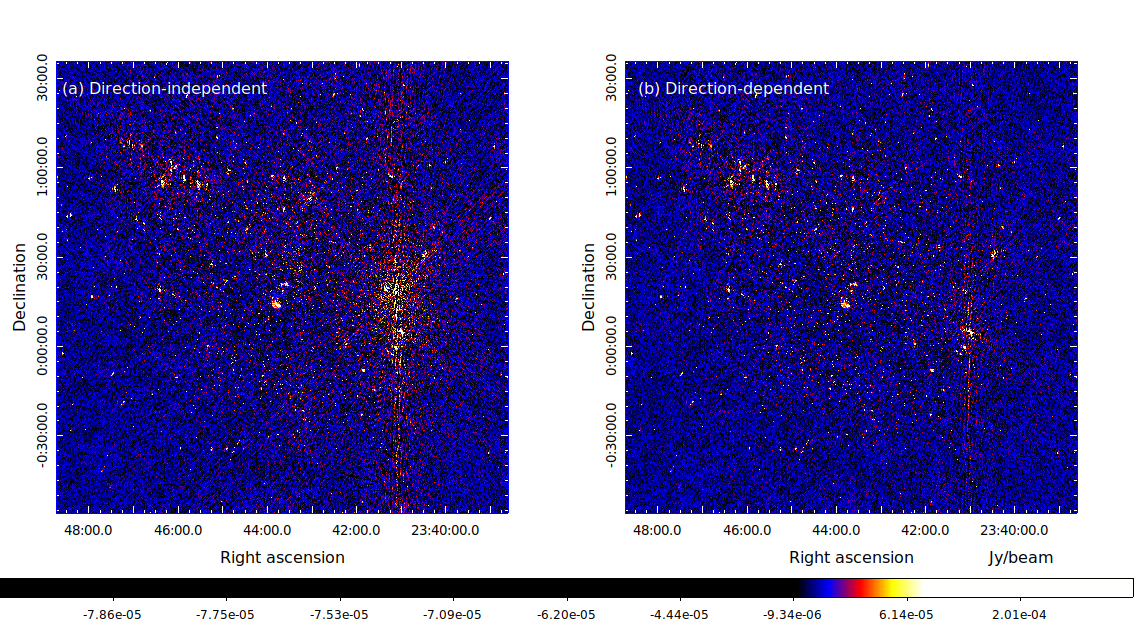}
\caption{Full MeerKAT image (1$^{\circ}$ field of view) of ZwCl2341.1+0000. Left image is direction-independent and right image is direction-dependent image. In both images, the beam size is the same as Figure \ref{dd_labled_marked}. The rms in the left image is 20 $\mu$Jy beam$^{-1}$ and right image is 14 $\mu$Jy beam$^{-1}$.}
\label{zwcl_dd_calibration}        
\end{figure*}
\end{center}

 \begin{center}
 \begin{figure*}
     \centering
     \begin{subfigure}[t]{0.75\textwidth}
         \includegraphics[width=1\textwidth]{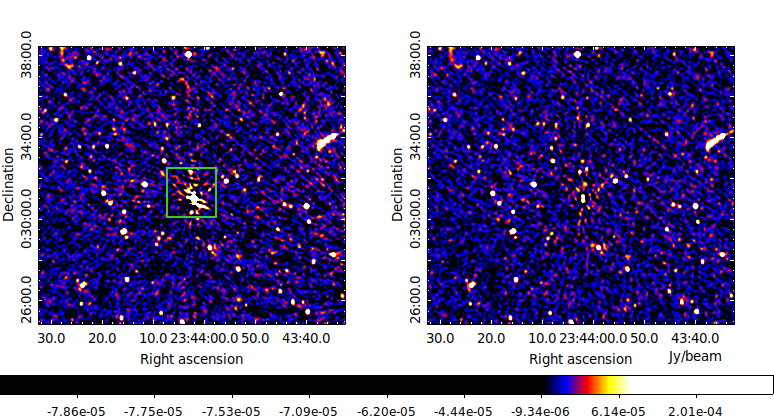}
         \caption{}
         \label{rfidtest_xaxis1}
     \end{subfigure}
     \begin{subfigure}[t]{0.75\textwidth}
         \includegraphics[width=1\textwidth]{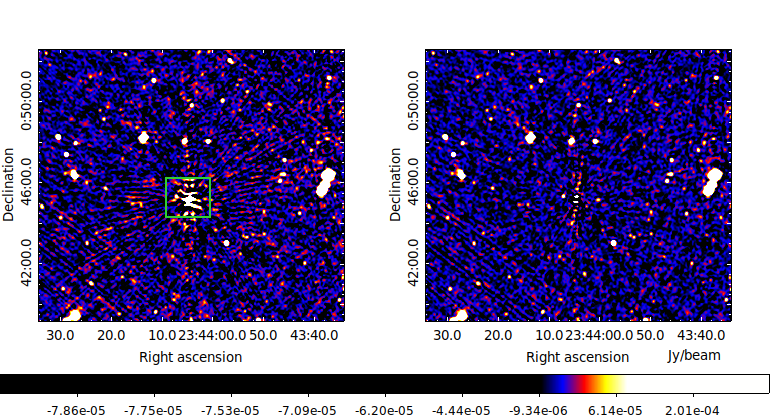}
         \caption{}
         \label{rfidtest_yaxis2}
         \end{subfigure}
     \begin{subfigure}{0.75\textwidth}
         \includegraphics[width=1\textwidth]{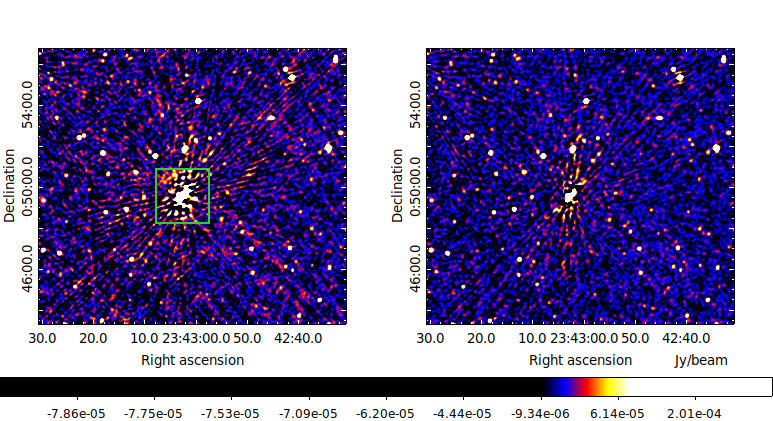}
         \caption{}
         \label{rfidtest_xaxis1}
     \end{subfigure}
    \end{figure*}
    \begin{figure*}\ContinuedFloat
    \centering
     \begin{subfigure}[t]{0.75\textwidth}
         \includegraphics[width=1\textwidth]{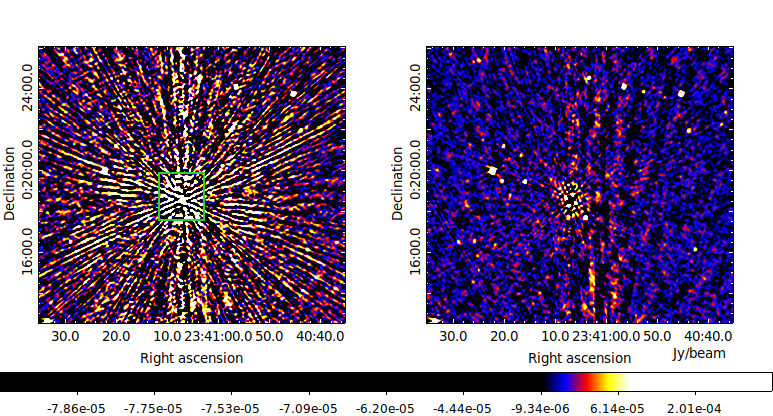}
         \caption{}
         \label{rfidtest_yaxis2}
         \end{subfigure}
      \caption{ZwCl 2331.1+0000: cutouts of four point sources towards direction-dependent calibration applied and peeled from the visibility. Sources are arranged from their distance from the phase-centre. We marked every source in box. Left cutout (from Figure \ref{zwcl_dd_calibration}(a)) is direction-independent and right cutout (from Figure \ref{zwcl_dd_calibration}(b)) is after direction-dependent calibration. In each cutout, we calculated the local rms, minimum pixel and sum of total negative component values inside the 8$'$ box size around the marked sources. These values are listed in Table \ref{DD_stat}.}
     \label{zwcl_dde_compt_src}        
 \end{figure*}
 \end{center}
 
\begin{center}
\begin{figure*}
\centering
\includegraphics[width=1\textwidth]{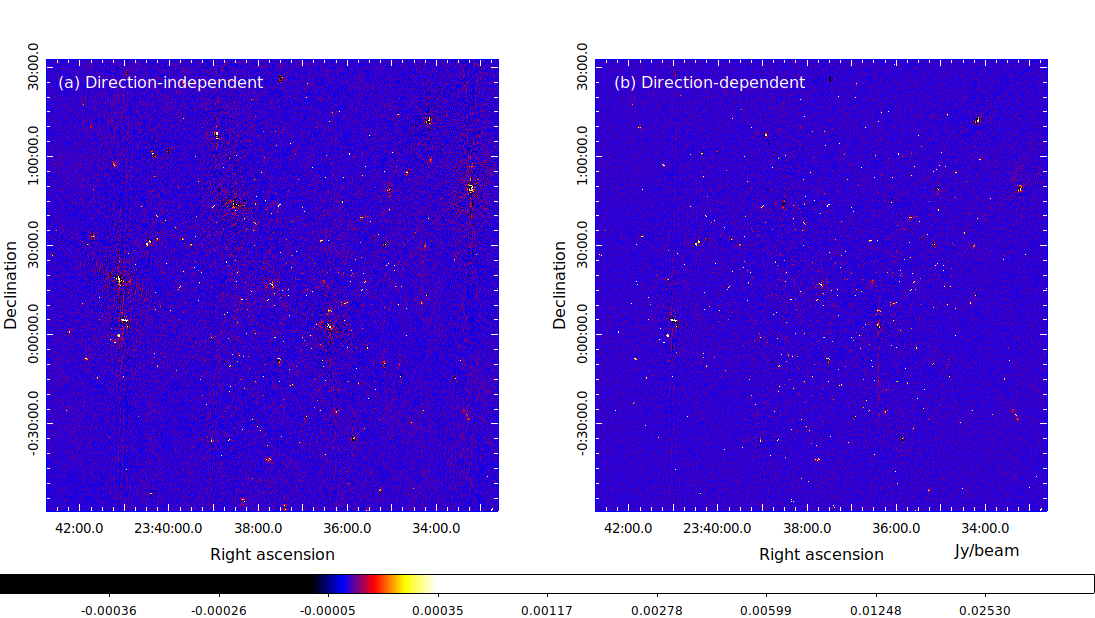}
\caption{Full MeerKAT image (1$^{\circ}$ field of view) of A2631. Left image is direction-independent and right image is direction-dependent image. In both images, beam size is the same as Figure \ref{dd_labled_marked}. The rms in the left image is 32 $\mu$Jy beam$^{-1}$ and right image is 13 $\mu$Jy beam$^{-1}$.}
\label{A2631_dd_calibration}        
\end{figure*}
\end{center}

 \begin{center}
 \begin{figure*}
     \centering
     \begin{subfigure}[t]{0.75\textwidth}
         \includegraphics[width=1\textwidth]{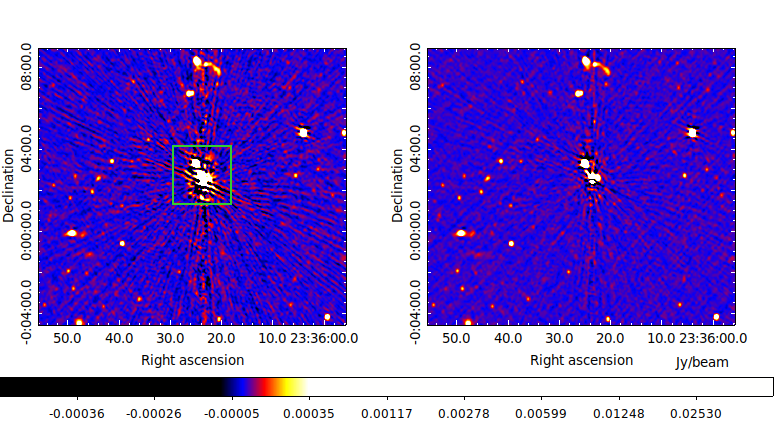}
         \caption{}
         \label{rfidtest_xaxis1}
     \end{subfigure}
     \begin{subfigure}[t]{0.75\textwidth}
         \includegraphics[width=1\textwidth]{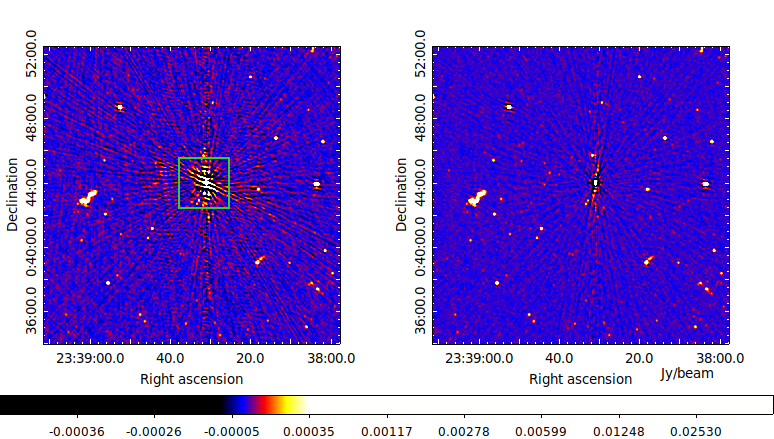}
         \caption{}
         \label{rfidtest_yaxis2}
         \end{subfigure}
     \begin{subfigure}[t]{0.75\textwidth}
         \includegraphics[width=1\textwidth]{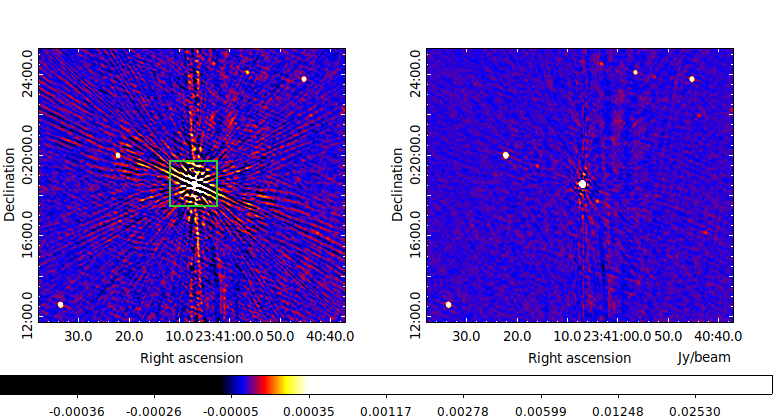}
         \caption{}
         \label{rfidtest_xaxis1}
     \end{subfigure}
      \caption{A2631: cutouts of three point sources towards direction-dependent calibration applied and peeled from the visibility. Sources are arranged from their distance from phase-centre. We marked every source by box. Left cutout (from Figure \ref{A2631_dd_calibration}(a)) is direction-independent and right cutout (from Figure \ref{A2631_dd_calibration}(b)) is after direction-dependent calibration. In each cutout, we calculated the local rms, minimum pixel and sum of the total negative component values inside the 8$'$ box size around the marked sources. These values are listed in Table \ref{DD_stat}.}
     \label{A2631_dde_compt_src}        
 \end{figure*}
 \end{center}

\begin{center}
\begin{figure*}
\centering
\includegraphics[width=0.5\textwidth]{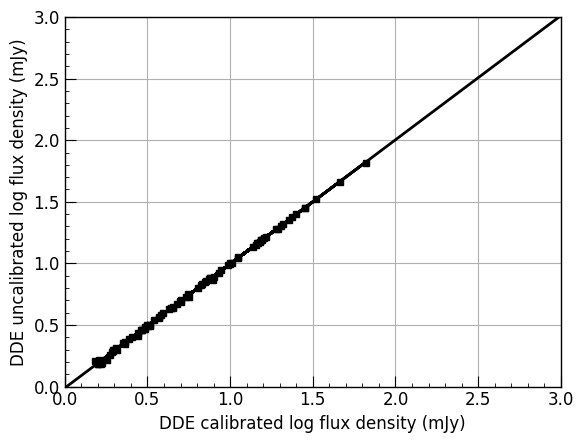}
\caption{Flux densities comparison of discrete point sources before and after direction-dependent calibration for both A2631 and ZwCl 2341.1+0000.}
\label{dd_calibration_compare}        
\end{figure*}
\end{center}

\begin{table*}
\caption{Local rms, minimum of pixel and sum of total negative component values calculated around peeled sources for two sets of cutouts of direction-independent (DI) and direction-dependent (DD).}
\label{DD_stat}
\begin{tabular}{cccccccc}
\hline
\hline
\multicolumn{1}{c}{Clusters} & \multicolumn{2}{c}{RMS ($\mu$Jy beam$^{-1}$)} & \multicolumn{2}{c}{ Min. of pixel (mJy beam$^{-1}$)} & \multicolumn{2}{c}{ Sum of negative (Jy beam$^{-1}$)}\\
 &                     DI& DD      & DI& DD &     DI&     DD  \\
\hline     
ZwCl2341.1+0000  &     37& 16     & -1.26 &-0.091  &-1.17&  -0.63 \\
     &                 23& 17     & -0.40&-0.20   &-0.77&  -0.64 \\
     &                 37& 18     & -1.11&-1.57   &-0.98 & -0.72  \\
     &                 38& 24     & -1.33&-0.61   &-0.99 & -0.65  \\
     &                 201& 30     & -8.23&-60.0   &-4.13 & -2.53  \\
A2631            &     113& 41     & -2.21&-1.84    &-2.22 & -1.14 \\
     &                 147& 29     & -3.75& -1.56   &-2.65 & -0.93 \\
     &                 183& 21      & -4.02& -0.28   &-3.29 &  -0.68 \\
                      \hline    
\end{tabular}
\end{table*}


\begin{table*}
\caption{List of point sources subtracted from the ZwCl 2341.1+0000 data.}
\label{pt_list}
\begin{tabular}{cccccccc}
\hline
\hline
Point source & RA & Declination & Flux density& Flux density error & Peak flux & Peak flux error\\
             & H~:M~:S~ & D~:M~:S~ & mJy & mJy & mJy beam$^{-1}$ & mJy beam$^{-1}$\\
\hline         
A1 & 23:43:39.4 & +0:18:45.0 &  15.59 & 1.36 & 4.65 & 0.32  \\
A2 & 23:43:40.8 & +0:20:30.0 &  0.89 & 0.01 & 0.82 & 0.01    \\
A3 & 23:43:40.3 & +0:20:54.0 &  3.67 & 0.14 & 2.31 & 0.06   \\
A4 & 23:43:39.7 & +0:17:55.5 &  0.41 & 0.01 & 0.36 & 0.00   \\
A5 & 23:43:37.5 & +0:17:28.5 &  0.30 & 0.00 & 0.31 & 0.00   \\
A6 & 23:43:47.3 & +0:16:55.5 &  0.41 & 0.01 & 0.39 & 0.01   \\
A7 & 23:43:46.0 & +0:15:34.5 &  1.29 & 0.02 & 1.16 & 0.01    \\
A8 & 23:43:50.1 & +0:22:40.5 &  7.20 & 0.07 & 6.92 & 0.04   \\
A9 & 23:43:58.6 & +0:14:59.9 &  1.03 & 0.01 & 0.99 & 0.01   \\
A10 & 23:43:56.0 & +0:18:31.4 & 0.24 & 0.01 & 0.23 & 0.01   \\
A11 & 23:43:34.7 & +0:20:37.5 & 0.64 & 0.01 & 0.57 & 0.01   \\
A12 & 23:43:41.4 & +0:18:27.0 & 0.17 & 0.03 & 0.14 & 0.01    \\
A13 & 23:43:43.9 & +0:20:39.0 & 0.40 & 0.01 & 0.31 & 0.01   \\
A14 & 23:43:23.8 & +0:16:11.9 & 0.70 & 0.01 & 0.67 & 0.01   \\
A15 & 23:43:48.5 & +0:15:09.0 & 0.93 & 0.06 & 0.66 & 0.03   \\
A16 & 23:43:49.8 & +0:15:07.5 & 0.76 & 0.05 & 0.39 & 0.02  \\
A17 & 23:43:48.8 & +0:15:31.5 & 0.96 & 0.02 & 0.50 & 0.01   \\
A18 & 23:43:45.9 & +0:16:09.0 & 2.91 & 0.19 & 0.98 & 0.05  \\
A19 & 23:43:34.5 & +0:15:31.5 & 0.20 & 0.01 & 0.17 & 0.00  \\
A20 & 23:43:34.5 & +0:17:00.0 & 0.23 & 0.03 & 0.18 & 0.01  \\
A21 & 23:43:25.0 & +0:16:53.9 & 0.27 & 0.01 & 0.27 & 0.01  \\
A22 & 23:43:42.2 & +0:15:42.0 & 0.19 & 0.01 & 0.16 & 0.00   \\
A23 & 23:43:36.3 & +0:20:51.0 & 0.40 & 0.01 & 0.26 & 0.00  \\
A24 & 23:43:33.9 & +0:20:00.0 & 0.30 & 0.01 & 0.22 & 0.00  \\
A25 & 23:43:32.9 & +0:14:28.5 & 0.26 & 0.00 & 0.23 & 0.00   \\
A26 & 23:43:37.3 & +0:20:38.5 & 0.17 & 0.01 & 0.14 & 0.00  \\
A27 & 23:43:37.0 & +0:21:41.4 & 0.28 & 0.00 & 0.28 & 0.00  \\
A28 & 23:43:48.0 & +0:14:00.8 & 0.44 & 0.00 & 0.32 & 0.00  \\

\hline    
\end{tabular}
\end{table*}




\end{document}